\documentclass[twocolumn]{aastex6}
\pdfoutput=1 
\usepackage{amsmath,amstext}
\usepackage{amssymb}
\usepackage{rotating}
\usepackage[T1]{fontenc}
\usepackage{apjfonts} 
\usepackage[figure,figure*]{hypcap}
\usepackage{gensymb}
\usepackage{textcomp}
\usepackage{verbatim} 
\usepackage{tablefootnote}
\usepackage{cleveref}
\usepackage{float}

\usepackage{amsmath}

\newcommand{\RN}[1]{%
  \textup{\uppercase\expandafter{\romannumeral#1}}%
}
\shorttitle{AASTeX 6 Template}
\shortauthors{LU et al.}

\begin{document}

\title{A Statistical study of the magnetic imprints of X-class solar flares}

\author{Zekun Lu\altaffilmark{1}, Weiguang Cao\altaffilmark{1},Gaoxiang Jin\altaffilmark{1}, Yining Zhang\altaffilmark{1}, Mingde Ding\altaffilmark{1,2}, Yang Guo\altaffilmark{1,2}}
\email{dmd@nju.edu.cn}
\email{guoyang@nju.edu.cn}
\altaffiltext{1}{School of Astronomy and Space Science, Nanjing University, Nanjing 210023, China;{ \color{blue} dmd@nju.edu.cn}}
\altaffiltext{2}{Key Laboratory for Modern Astronomy and Astrophysics (Nanjing University), Ministry of Education, Nanjing 210023, China; { \color{blue}guoyang@nju.edu.cn}
}

\begin{abstract}
Magnetic imprints, the rapid and irreversible evolution of photospheric magnetic fields as a feedback from flares in the corona, have been confirmed by many previous studies. These studies showed that the horizontal field will permanently increase {at} the polarity inversion line (PIL) after eruptions, indicating that a more horizontal geometry of photospheric magnetic field is produced. In this study, we analyze {20 X-class flares} since the launch of the Solar Dynamics Observatory (SDO) in {15 active regions (ARs)} with heliographic angle no greater than $45^\circ$. {We observe clear magnetic imprints in 16 flares, whereas 4 flares are exceptional.} The imprint regions of the horizontal field are located not only at the PIL but also at sunspot penumbra with strong vertical fields. Making use of the observed mass and speed of the corresponding coronal mass ejections (CMEs) , we find that the CMEs with larger momentums are associated with stronger magnetic imprints. {Furthermore, a linear relationship, with a Kendall's Tau-b coefficient 0.54, between the CME momentum and the change of Lorentz force is revealed. Based on that, we quantify the back reaction time to be {$\sim$70 s}, with a 90\% confidence interval from about 50 s to 90 s.}
\end{abstract}

\section{Introduction}
\begin{center}
\begin{figure*}[t]
\label{fig:fig1}
\centering
\includegraphics[width=18cm]{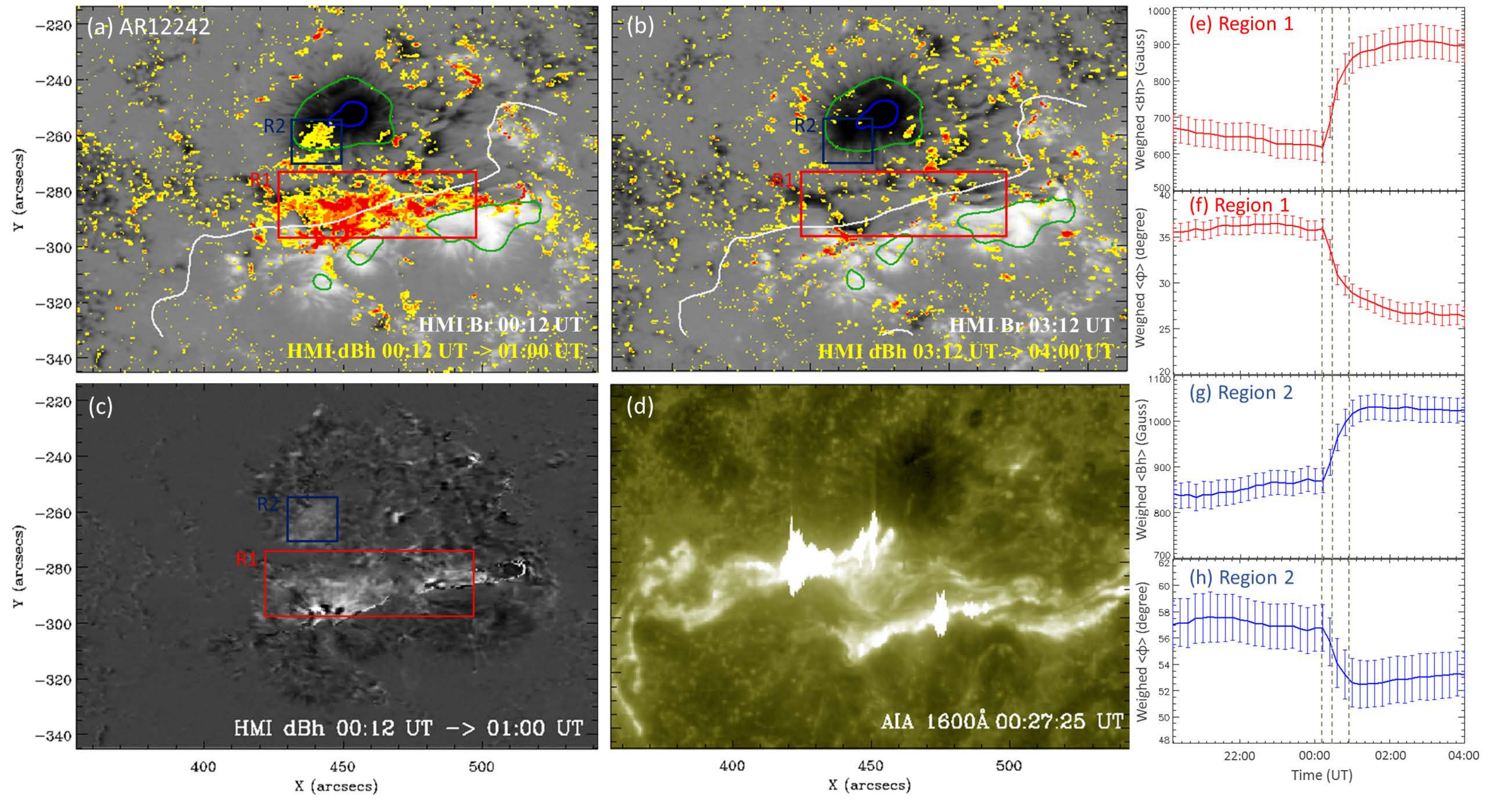}
\caption{X1.8 flare on 2014 December 20. {Panels (a) and (b) show the change in the magnitude of the horizontal field (the color map), $|\delta B_h|$, with a threshold 100 G, during and after the solar flare. The areas with a change, $|\delta B_h|$, greater than 100 G, 200 G and 300 G are marked in yellow, orange and red respectively.} In panel (a), we show the difference map of $B_h$ in flare stage (from 00:12 UT to 01:00 UT). In panel (b), we plot the difference map of $B_h$ in post-flare stage (from 03:12 UT to 04:00 UT, three hours after flare stage). { The background gray map shows the radial magnetic field $B_r$ scaled between $\pm$2000G, where $|B_r|$= 1000 G and 2000 G are contoured using green line and blue line respectively. The white line shows the smoothed PIL, where $B_r$= 0 G.} The red box encloses the region (Region 1) with flare-associated enhancement of $B_h$ at the Polarity Inversion Line (PIL); { the blue box encloses} the region (Region 2) with such change at penumbra. Panel (c) is the difference map of horizontal field, with contrast strengthened. Panel (d) is an AIA 1600\ \AA \ map around flare peak time. In panels (e)--(h), we plot the eight-hours evolution of the weighted mean horizontal field and weighted mean magnetic inclination of { ROI (the colored areas inside Region 1 and Region 2)}, where gray dashed lines mark start, peak and end time of the flare. The vertical error bars are calculated through error transfer formula.}
\end{figure*}
\end{center}

Regarded as a rapid energy release process by virtue of magnetic reconnection in the solar corona \citep{Kopp1976}, {major solar flares} are usually associated with coronal mass ejections (CMEs). It is known that long-term evolution of photospheric magnetic field plays an essential role in storing the energy responsible for the flares. However, since the corona is much more tenuous than the photosphere, it has been thought for a long time that solar flares cannot influence the photospheric magnetic fields, especially in short-term evolution.

{The first observational evidence} of rapid/irreversible flare-related change of the photospheric magnetic field was reported by \cite{Wang1992} and \cite{1994ApJ...424..436W}. {A strong and permanent increase of the magnetic shear at the polarity inversion line (PIL) was found.} Subsequently, \cite{Wang2002_0004-637X-576-1-497} found that the rapid permanent changes of photospheric magnetic fields associated with six X-class flares. A survey covering fifteen major X-class flares undertaken by \cite{Sudol2005_0004-637X-635-1-647} concluded that the line-of-sight (LOS) magnetic field always changes during X-class flares. The above studies were based on magnetic fields observed by ground-based telescopes.

Nevertheless, ground-based observations have rarely provided time series of magnetograms covering the flare time with a sufficiently high cadence. On the other hand, previous studies are mostly based on the LOS magnetic field data that are actually a combination of both the horizontal component and the radial one. This makes the interpretation of the results more challenging. {This intrinsic limitation was solved with high-cadence photospheric vector magnetic field data { \citep{Hoeksema-2014SoPh..289.3483H}} acquired by the Helioseismic and Magnetic Imager {(HMI; \citealt{Scherrer-et-al-2012})} , which was successfully launched aboard the Solar Dynamics Observatory {(SDO; \citealt{Pesnell-et-al-2012})}.} Since then, many more flare events showing rapid and irreversible  imprints on photospheric magnetic field were found (\citealt{Wangshuo2012a_2041-8205-745-2-L17}, \citealt{Wangshuo2012b_2041-8205-757-1-L5}; \citealt{Xudong2012_0004-637X-748-2-77}; \citealt{Petrie2012_0004-637X-759-1-50}, \citealt{Petrie2013}; \citealt{song2016relationship}). {\cite{ye2016irreversible} observed the rapid/permanent change of photospheric magnetic field in a circular-ribbon flare, which is associated with the magnetic reconnection of a three-dimensional magnetic null point (\citealt{Wang-Liu-2012ApJ-circular-ribbon-flares}; \citealt{Masson-2009ApJ...700..559M}, \citeyear{Masson-2017A&A...604A..76M}; \citealt{Yang-2015ApJ...806..171Y}; \citealt{Liu-C.-2015ApJ...812L..19L}).} \cite{Xudong2017_0004-637X-839-1-67} studied on the magnetic imprints for a sample of X-class flares using the recently-released SDO/HMI high-cadence vector magnetograms. The main findings of previous studies are that the horizontal magnetic field tends to increase at the PIL, indicating that vector magnetic field tends to be more horizontal after flare eruptions. Theoretically, the tether-cutting model \citep{Moore2001_0004-637X-552-2-833} and the coronal implosion model \citep{Hudson2000_1538-4357-531-1-L75} have been referred to explain this process. 

{Although previous observations have revealed many observational aspects of the imprints of solar flares}, the results are still not conclusive and thus more statistical studies are required. A statistical work for 18 flare events from 4 Active Regions (ARs) was conducted by \cite{Wangshuo2012b_2041-8205-757-1-L5}. \cite{Petrie2012_0004-637X-759-1-50} studied six major flare events in detail and \cite{Xudong2017_0004-637X-839-1-67} investigated nine X-class flares. {It should be mentioned that the events in these studies are overlapping and mainly belong to major eruptions from a small number of distinct active regions: AR 11158, 11166, 11283 and 11429.} Thus, to avoid selection bias, we conduct a comprehensive survey of all the X-class flares with heliographic angle no greater than $45\degree$ since SDO's launch. 

{As for the cause of the magnetic imprints, \cite{Fisher2012} considered the net Lorentz force acting on the upper solar atmosphere, and equate it (with an opposite direction) with the net Lorentz force acting on the interior of the Sun. Meanwhile, the downward momentum incurred by Lorentz force should equal to the upward momentum of the erupting plasma.} Based on this theory, \cite{Wangshuo2012b_2041-8205-757-1-L5} estimated the CME mass by assuming that the back reaction time is of order of 10 s. In this work, we investigate the relationship between CME momentum and change of the the Lorentz force, where we will elucidate that change of the Lorentz force can serve as an indicator of the strength of the magnetic imprint. Moreover, instead of {an empirical estimation}, we give a more exact evaluation of the reaction time, which is an important parameter in understanding the reaction process.

\begin{center}
  \begin{table*}[t]
	\centering
    \caption{List of 20 flare events from 15 ARs with heliographic angle no greater than 45$\degree$}
  \label{tab:flarelist}
  \begin{tabular}{ c  c  c  c  c c c c c c c c c}
\hline
     NOAA &Date       &GOES  &Location  &Start &Stop  &Peak  &Category\footnote{{ The category is defined in Section \ref{sec:results}.}}      &$\langle\ \delta B_h\ \rangle$\footnote{The weighted mean change of horizontal field $B_h$ and inclination angle $\phi$ are defined by Equations (\ref{eq:WeightedBh}) and (\ref{eq:Weightedtheta}).}  &$\langle\ \delta \phi\ \rangle$     &CME Momentum\footnote{The CME momentums are computed by multiplying the CME mass and CME speed.} \\
       AR &           &Class& &(UT)  &(UT)  &(UT)   &    & (Gauss) & ($\degree$) &($10^{17} g\cdot km\cdot  s^{-1}$) \\\hline \hline
 
    11158 &2011.02.15 &X 2.2  & S21W28 &01:44 &02:06 &01:56 &Type \RN{1}   &286$\pm$59 &6.23$\pm$0.47 &28.77\\
    11166 &2011.03.09 &X 1.5  & N11W15 &23:13 &23:29 &23:23 &Type \RN{1}   &258$\pm$73 &2.31$\pm$1.13   &0.40\\
    11283 &2011.09.06 &X 2.1  & N14W18 &22:12 &22:24 &22:20 &Type \RN{1}   &339$\pm$53 &3.70$\pm$0.50  &86.25 \\
    11283 &2011.09.07 &X 1.8  & N14W32 &22:32 &22:44 &22:38 &Type \RN{1}   &260$\pm$66 &7.84$\pm$0.64  &8.71 \\
    11302 &2011.09.24 &X 1.9  & N13E45 &09:21 &09:48 &09:40 &Type \RN{1}   &197$\pm$99 &10.49$\pm$1.65 &54.21\\
    11429 &2012.03.07 &X 5.4  & N17E27 &00:02 &00:40 &00:24 &Type \RN{1}   &216$\pm$73 &9.30$\pm$0.35  &375.76\\
    11430 &2012.03.07 &X 1.3  & N22E12 &01:05 &01:23 &01:14 &Type \RN{1}   &170$\pm$95 &5.75$\pm$0.76   & - -\\
    11890 &2013.11.05 &X 3.3  & S13E44 &22:07 &22:15 &22:12 &Type \RN{1}   &454$\pm$121&3.51$\pm$1.21  &26.98\\
    11890 &2013.11.08 &X 1.1  & S14E15 &04:20 &04:29 &04:26 &Type \RN{1}   &387$\pm$56 &7.55$\pm$0.63   & - -\\
    11890 &2013.11.10 &X 1.1  & S14W13 &05:08 &05:18 &05:14 &Type \RN{1}   &285$\pm$57 &4.10$\pm$0.60  &15.69\\

    12017 &2014.03.29 &X 1.0  & N11W32 &17:35 &17:54 &17:48 &Type \RN{1}   &223$\pm$63 &10.29$\pm$0.75 &26.40\\
    12205 &2014.11.07 &X 1.6  & N17E40 &16:53 &17:34 &17:26 &Type \RN{1}-R1&323$\pm$78 &8.79$\pm$0.82  &95.40\\
          &           &       &        &      &      &      &Type \RN{1}-R2&278$\pm$75 &7.40$\pm$0.83     &\\    
    12297 &2015.03.11 &X 2.2  & S17E21 &16:11 &16:29 &16:22 &Type \RN{1}   &309$\pm$102&4.00$\pm$1.04     &2.64\\
    11520 &2012.07.12 &X 1.4  & N13W15 &15:37 &17:30 &16:49 &Type \RN{2}-R1&212$\pm$33 &5.73$\pm$0.81 &61.06\\ 
          &           &       &        &      &      &      &Type \RN{2}-R2&215$\pm$88 &1.17$\pm$0.46  &\\
    12158 &2014.09.10 &X 1.6  & N15E02 &17:21 &17:45 &17:45 &Type \RN{2}-R1&161$\pm$22 &5.80$\pm$1.56 &- -\\ 
          &           &       &        &      &      &      &Type \RN{2}-R2&167$\pm$15 &3.67$\pm$0.96 & \\
	12242 &2014.12.20 &X 1.8  & S21W24 &00:11 &00:55 &00:28 &Type \RN{2}-R1&261$\pm$58 &9.09$\pm$0.35  &182.60\\
          &           &       &        &      &      &      &Type \RN{2}-R2&161$\pm$39 &4.25$\pm$0.60    &\\
11944 &2014.01.07 &X 1.2 & S12W08  &18:04 &18:58 &18:32 &Type \RN{3}      
& - -             & - -     &402.6  \\
    12192 &2014.10.22 &X 1.6  & S14E13 &14:02 &14:50 &14:28 &Type \RN{3}   &- - &- -   &0.50\\
    12192 &2014.10.24 &X 3.1 & S16W21  &21:07 &22:13 &21:44 &Type \RN{3}   &- -      
&- -                   &0.0552      \\
    12192 &2014.10.26 &X 2.0  & S18W40 &10:04 &11:18 &10:56 &Type \RN{3}   &- - &- - &- -       \\
    \hline

    \end{tabular}
  \end{table*}

\end{center}

Our motivations for this study are fourfold: (1) to investigate the general properties of flares' imprints on the photosphere; (2) to figure out the location of the imprints { and categorize them}; (3) to quantitatively describe the change of magnetic field including horizontal field change and inclination angle change; and (4) to further study the relationship between CMEs and the imprints on the photosphere. In Section \ref{sec:observation}, we describe the observations and the data processing procedure for 20 events with AR 12242 as an example. The statistical results for the magnetic field change are presented in Section \ref{sec:results}, together with the relationship between the CME momentum and the change of Lorentz force. We finally make a discussion and summary in Section \ref{sec:conclusion}.
\label{sec:intro}

\section{Observation and Data Reduction}
\label{sec:observation}

For the events occurring near the solar limb, when compared with those near the disk center, observational noise level is typically higher; moreover, it is harder to spatially resolve small structures of solar flares since a pixel covers a larger area of the solar surface. Therefore, to guarantee the reliability of our study, we only study {the near-disk center X-class} flares with heliographic angle no greater than 45\degree. We scrutinize the whole list of X-class flares\footnote{\href{ftp://ftp.ngdc.noaa.gov/STP/space-weather/solar-data/solar-features/solar-flares/x-rays/goes/xrs/}{\color{blue}ftp://ftp.ngdc.noaa.gov/STP/space-weather/solar-data/solar-features/solar-flares/x-rays/goes/xrs/} }\footnote{\href{https://www.solarmonitor.org/}{\color{blue}https://www.solarmonitor.org/}}\footnote{\href{https://helioviewer.org/}{\color{blue}https://helioviewer.org/}}, from SDO's launch in 2010 to 2017 June 28, and {select 20 flare events from 15 ARs} (listed in Table \ref{tab:flarelist}). The GOES X-ray classes of the sample flares ranges from X1.0 to X5.4, which is to date the largest sample for studying magnetic imprints of X-class flares since SDO's launch. {Most importantly, we do not make other subjective selection criteria than setting a threshold for the flare class and heliographic angle of the flare site, which avoids bias in the statistical results.}

HMI and Atmospheric Imaging Assembly (AIA; \citealt{Lemen2012}) on board SDO provide full-disk vector magnetic fields and Extreme Ultra-Violet (EUV) images, respectively, with high spatial and temporal resolutions. More specifically, we trace the morphology of flare ribbons using the AIA images at wavelength 1600\ \AA, with a cadence of 24 s (aia.lev1\_uv\_24s\footnote{\href{http://jsoc.stanford.edu/ajax/lookdata.html?ds=aia.lev1_uv_24s}{\color{blue}http://jsoc.stanford.edu/ajax/lookdata.html?ds=aia.lev1\_uv\_24s}}). HMI acquires full-disk photospheric magnetograms with a pixel size of $0.5''$ at a cadence of 12 minutes. The magnetic field is inverted from the Stokes parameters at six wavelengths distributed around the photospheric FeI 6173\ \AA \ line\footnote{\href{http://jsoc.stanford.edu/jsocwiki/VectorMagneticField}{\color{blue}http://jsoc.stanford.edu/jsocwiki/VectorMagneticField}}. The 180\degree \ azimuthal ambiguity is resolved through the "minimum energy" method (\citealt{Metcalf1994}; \citealt{Leka2009}). The noise level of the LOS field measurement is about 10 G; the noise level of the transverse field is of the order of 100 G \citep{Liu2012}. In this work, we use the vector magnetogram Spaceweather HMI Active Region Patch (SHARP), where the ARs are automatically identified and extracted. { We obtained the three components of the vector magnetic fields, $B_r, B_p, B_t$ from the deprojected maps provided in the cylindrical equal area (CEA) coordinate system (hmi.sharp\_cea\_720s\footnote{\href{http://jsoc.stanford.edu/ajax/lookdata.html?ds=hmi.sharp_cea_720s}{\color{blue}http://jsoc.stanford.edu/ajax/lookdata.html?ds=hmi.sharp\_cea\_720s}}; \citealt{Bobra2014}).} 

As mentioned above, solar flares' imprints on the photosphere are mainly reflected by the irreversible increase of horizontal magnetic field, $B_h=\sqrt[]{B_p^2+B_t^2}$, {at the PIL}. Thus, for all events under study, we merely focus on the regions with magnetic field changes that meet two criteria:
\begin{enumerate}
\item A significant increase of $B_h$ is detected during the solar flare. To avoid unreal enhancement of $B_h$ arising from noise, we set a threshold of increase, 100 G, the contour of which is marked in yellow.
\item The increase of $B_h$ is temporally associated with solar flares. To get rid of non-flare-associated field changes, such as emerging flux or moving magnetic features (See \citealt{Hermance2005-0004-637X-635-1-659}; \citealt{Iida2012-0004-637X-752-2-149}), {we also calculate the change of $B_h$ in the post-flare stage, defined at an equal duration interval starting three hours after the start time of the flare.} If $B_h$ increases during the flare but does not increase after the flare, the increase is regarded to be associated with solar flares. The boundary of flare-associated increase of $B_h$ is marked by red or blue boxes.
\end{enumerate} Therefore, we define the regions of interest (ROI) as the regions that meet both the two criteria above, visually, the colored areas inside the boxes.

\begin{center}
\begin{figure*}[p!]
\label{fig:map1}
\centering
\includegraphics[width=13.5cm]{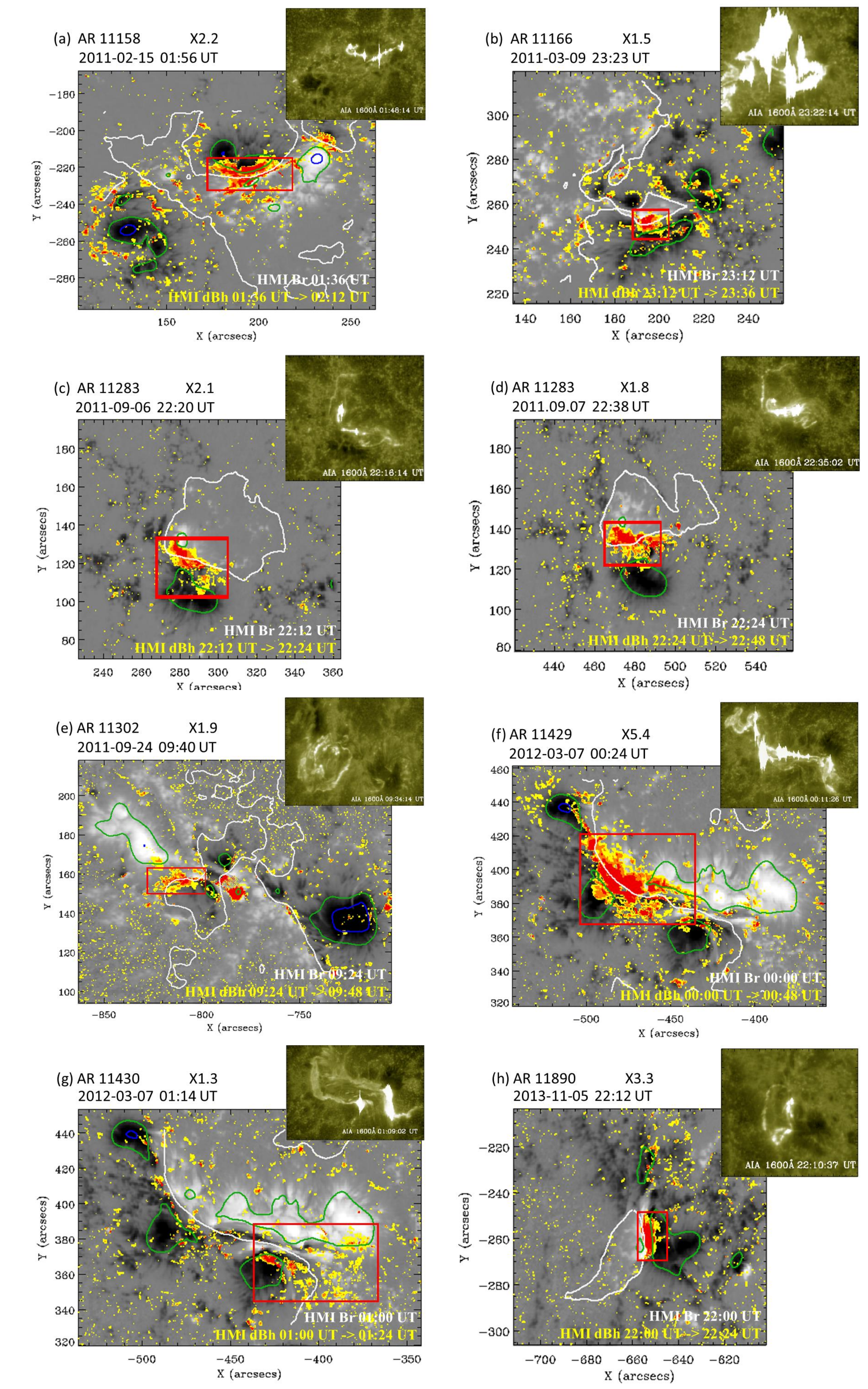}
\caption{Overview of magnetic imprints in eight X-class flares with heliographic angle no greater than 45\degree. The notation in each panel is similar to that in Figure \ref{fig:fig1}(a). The $B_r$ maps are all scaled between $\pm$ 2000 G. All the eight events in panels (a)--(h) are categorized as type \RN{1} in Section \ref{sec:results}.}
\end{figure*}
\end{center}

\begin{center}
\begin{figure*}[p!]
\label{fig:map2}
\centering
\includegraphics[width=13.5cm]{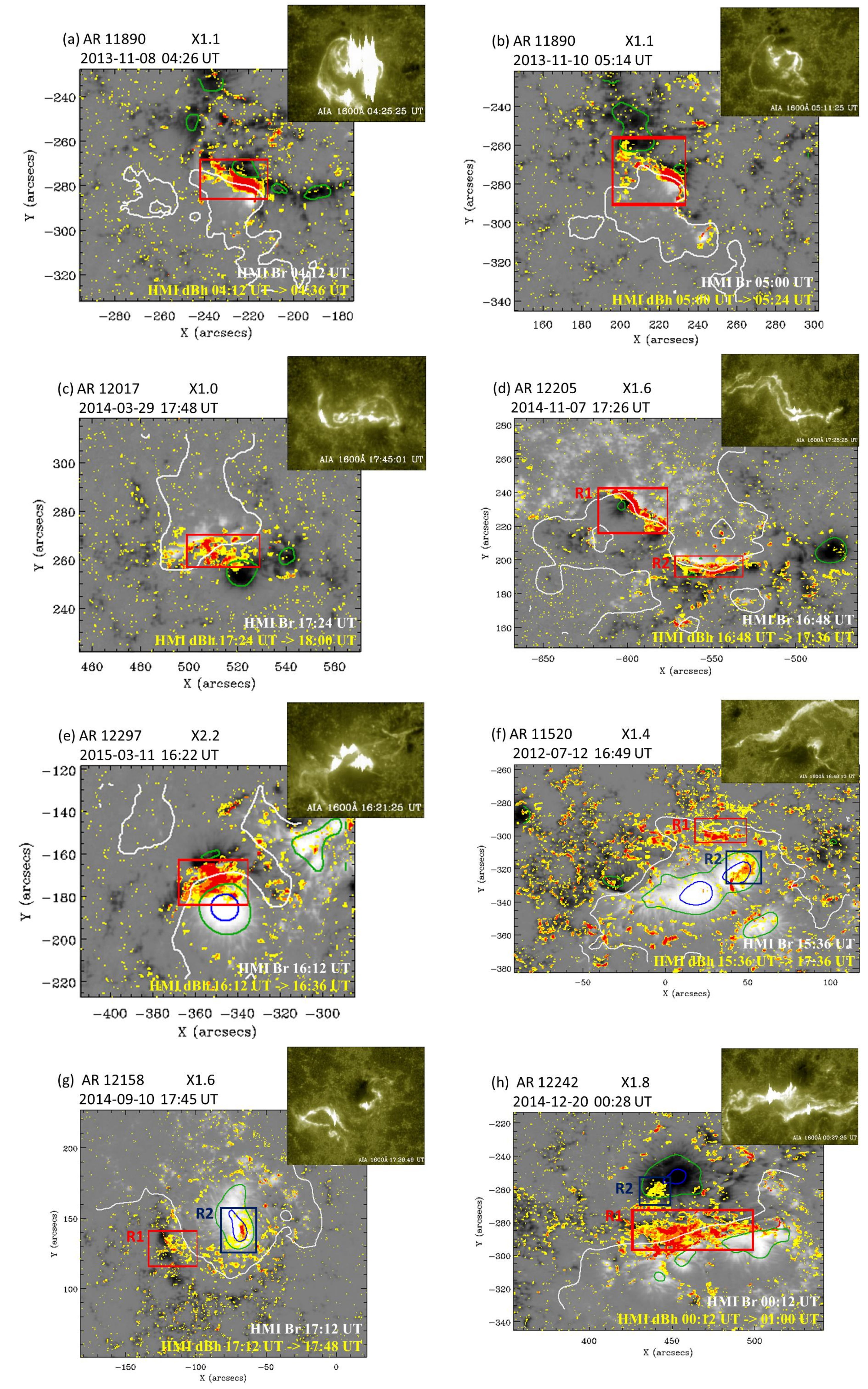}
\caption{Overview of magnetic imprints in eight X-class flares with heliographic angle no greater than 45\degree. Notation in each panel is similar to that in Figure \ref{fig:fig1}(a). The $B_r$ maps are all scaled between $\pm$ 2000 G. The five events in panels (a)--(e) are categorized as type \RN{1} in Section \ref{sec:results}. The three events in panels (f)--(h) are categorized as type \RN{2}.}
\end{figure*}
\end{center}

\begin{center}
\begin{figure*}[t!]
\label{fig:map3}
\centering
\includegraphics[width=1.0\textwidth]{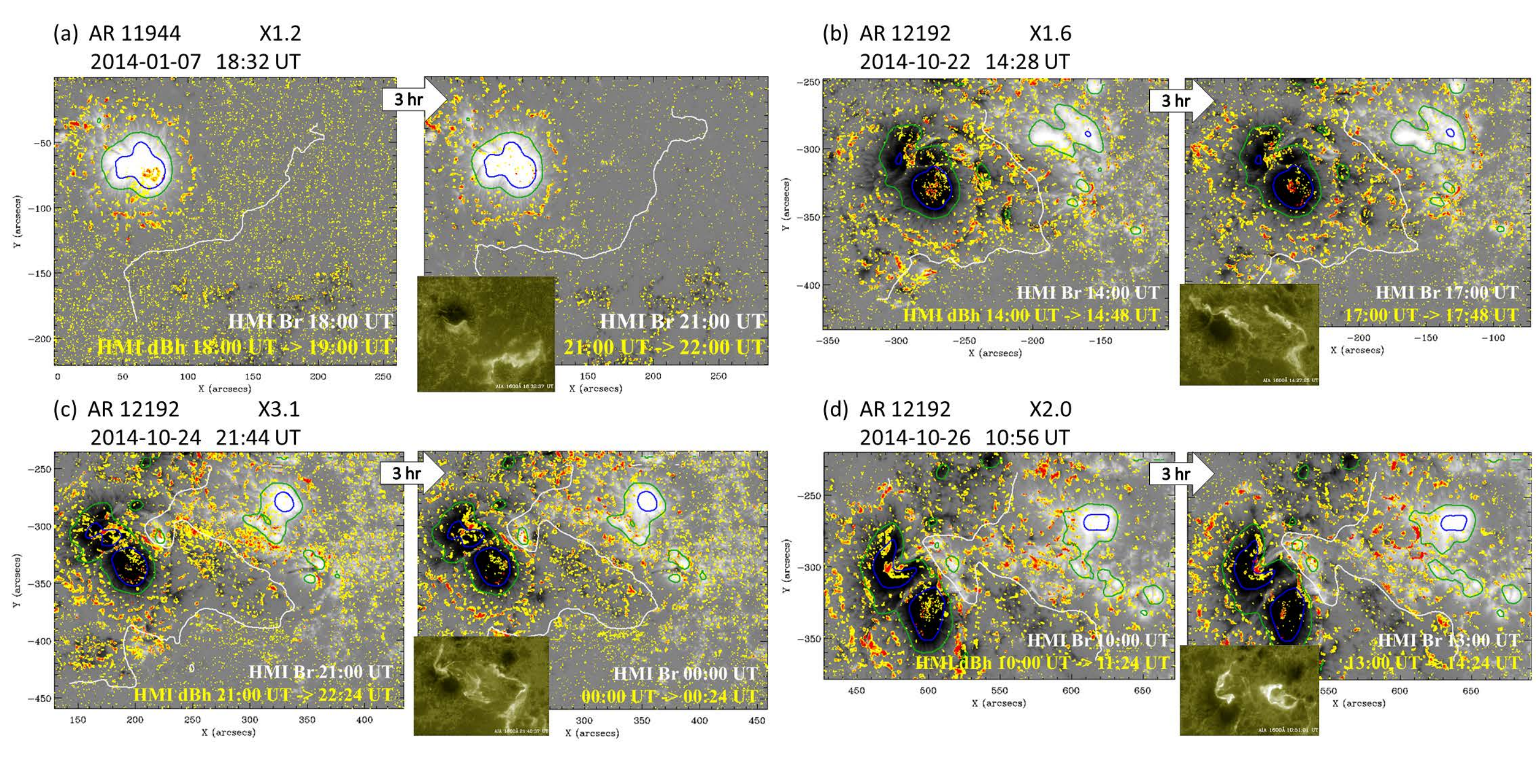}
\caption{Overview of magnetic imprints in four X-class flares with heliographic angle no greater than 45\degree. Notation in each panel is similar to those in Figure \ref{fig:fig1}(a), (b) and (d), where the left part shows the difference map of $B_h$ during flare time and the right part shows that in post-flare time (3 hr later). {The $B_r$ maps are all scaled between $\pm$ 2000 G.} The four events in panels (a)--(d) are categorized as type \RN{3} in Section \ref{sec:results}.}
\end{figure*}
\end{center}

To demonstrate the procedure of data processing, we take the analysis of the sample with the most complex imprint as an example: the X1.8 flare in AR 12242. Firstly, we scrutinize the evolution of the horizontal field and plot the difference map between flare start time and end times {(defined based on the recorded start and end times of GOES soft X-ray emission)}, as is seen in Figure \ref{fig:fig1}(c). {Secondly, as seen in Figure \ref{fig:fig1}(a) and (b), we plot the radial magnetic field, $B_r$, as the background, on which the regions with difference values, $\delta B_h$, greater than 100 G, 200G and 300 G, are marked in yellow, orange and red respectively. Panel (a) shows the difference map during the flare; while panel (b) shows the difference map three hours after the flare. Here, with a focus on the rapid/irreversible change of $B_h$ during flare time, we consider the post-flare change as the non-flare-associated change. Visually, we exclude the non-flare-associated change out of the difference map of $B_h$ in flare stage, then keep the remaining regions as flare-associated change, as marked by boxes. Therefore, the colored areas inside the boxes in Figure \ref{fig:fig1}(a) are ROI as defined aforementioned. We can see that, unlike the flare-associated change of $B_h$, the non-flare-associated change of $B_h$ is dispersed across the whole field of view.} Considering its low amplitude and small area, we regard the non-flare-associated change as mostly from noise or computational errors. In particular, we point out that the increase of horizontal field appears not only at the PIL, as marked by the red box, but also at sunspot penumbra region, where $B_r>1000$ G, as marked by the blue box.

{We will show in Section \ref{sec:results} that such double-region increase is not unique. There are other types of magnetic imprints. Thirdly, since we are interested in the spatial relationship between magnetic imprints and flare ribbons, we show the AIA image of flare region at 1600\ \AA \ in Figure \ref{fig:fig1}(d). We can find that, spatially, region R1 is located mainly between the two flare ribbons, whereas region R2 is located partially on the flare ribbon.} Note that the coordinates of SHARP maps have been transformed to helioprojective-Cartesian coordinates for comparison with AIA map. Finally, we plot the evolution of weighted mean change of the horizontal field and that of the inclination angle within the ROI as a function of time. The definitions of these parameters are as follows:

\begin{equation}
\label{eq:WeightedBh}
\langle\ B_h\ \rangle= \frac{\sum_{ROI} {|\mathbf{B}|}\cdot  B_h}{\sum_{ROI} \vert\mathbf{B}\vert}
\end{equation}

\begin{equation}
\label{eq:Weightedtheta}
\langle\ \phi\ \rangle= \frac{\sum_{ROI} {|\mathbf{B}|}\cdot \phi}{\sum_{ROI} \vert\mathbf{B}\vert}
\end{equation}

{Here, for better comparison with previous studies, we follow the definition of the magnetic inclination angle $\phi$ in \cite{Wangshuo2012a_2041-8205-745-2-L17}: $\phi=\arctan{\frac{B_r}{\sqrt{B_p^2+B_t^2}}}$.} As shown in Figure \ref{fig:fig1}(e)--(h), both $\langle B_h\rangle$ and $\langle \phi \rangle$ show a stepwise change in ~40 minutes, which is consistent with previous findings. {It is worth pointing out that because of the flare heating, the inversions of Stokes profiles are generally not reliable during the flare \citep{Hong-2018ApJ...857L...2H}, which further brings uncertainty to the magnetograms of HMI. Nevertheless, inversions of Stokes profiles are reliable before and after the flare, during which the quantities $B_h$ and $\phi$ almost keep constant. Thus, despite of the uncertainty during the flare, the irreversible/stepwise change of $B_h$ and $\phi$ seems realistic.} We also compare the region at the PIL and the region at the penumbra. { It is found that the changes of the horizontal field are similar in the two regions;} however, the change of the inclination angle in the PIL region is greater than that in the penumbra region. This result can be expected due to a stronger $B_r$ field in the latter. Note that the errors in $\langle B_h\rangle$ and $\langle \phi \rangle$ are calculated according to the error transfer formula.

To further reveal the origin of such an irreversible change of the magnetic field, we analyze the relationship between the CME momentum and the Lorentz force change as predicted by \cite{Fisher2012}. The change in the vertical component of the net Lorentz force acting on the volume below the photosphere is formulated as:
\begin{equation}
\label{eq:Lorentzforce}
\delta F_r=\frac{1}{8\pi}\int_{A_{ph}}dA\ (\delta B_r^2-\delta B_h^2).
\end{equation}
We note that, similar to previous results (\citealt{Wangshuo2012b_2041-8205-757-1-L5}, \citealt{song2016relationship}, \citealt{Xudong2017_0004-637X-839-1-67}), the vertical field $B_r$ shows no rapid change in the time duration of $8$ hours. Therefore, we omit the term $\delta B_r^2$ and calculate $\delta F_r$ in the ROI depending solely on the term $\delta B_h^2$, which means that the Lorentz force change can serve as a reasonable indicator of the strength of the magnetic imprints, namely:\begin{equation}\label{eq:LF2}
\delta F_r\footnote{For convenience, the $\delta F_r$ used in statistical study means the decrease of Lorentz force, which is the absolute value of that in Equations (\ref{eq:Lorentzforce}) and (\ref{eq:LF2}).}\approx\frac{1}{8\pi}\int_{A_{ROI}}dA\ (-\delta B_h^2).
\end{equation}  On the other hand, we estimate the CME momentum by directly multiplying the CME mass by the linear speed of the CME as observed by the SOHO satellite\footnote{\href{$https://cdaw.gsfc.nasa.gov/CME_list/$}{\color{blue}https://cdaw.gsfc.nasa.gov/CME\_list/}}. 

For each X-class flare, we pay attention to a time period of eight hours. We then apply the same procedure to all the 20 flare events as listed in Table \ref{tab:flarelist}. The detailed CME information and the change of the vertical Lorentz force are recorded in Table \ref{tab:CMElist}.

\begin{center}
  \begin{table*}[t]
	\centering
    \caption{List of 13 flare events with CME in Type \RN{1} and \RN{2}}
  \label{tab:CMElist}
  \begin{tabular}{ c  c  c  c  c c c c c}
  \hline
     NOAA &Date  &GOES   &CME Time\footnote{Information of the CME time is from the SOHO catalog.} &CME Speed &CME Mass &CME Momentum\footnote{The CME momentums are computed by multiplying the CME mass and CME speed.} &$\delta F_r$\\
     AR   &      &Class & (UT)  &  ($\mbox{km}\cdot \mbox{s}^{-1}$) & ($10^{15} \mbox{g}$)  &($10^{17} \mbox{g}\cdot \mbox{km}\cdot  \mbox{s}^{-1}$)& ($10^{22}\mbox{dyne}$)\\\hline 
       \hline
    11158 &2011.02.15 &X 2.2  & 02:24  &669   &4.3      &28.77&5.06\\
    11166 &2011.03.09 &X 1.5  & 23:05  &332   &0.12      &0.40&1.46\\
    11283 &2011.09.06 &X 2.1  & 23:05  &575   &15.0      &86.25&2.76\\
    11283 &2011.09.07 &X 1.8  & 23:05  &792   &1.1      &8.71&2.92\\
    11302 &2011.09.24 &X 1.9  & 09:48  &1936   &2.8      &54.21&1.33\\
    11429 &2012.03.07 &X 5.4  & 00:24  &2684   &14.0      &375.76&14.1\\
    11520 &2012.07.12 &X 1.4  & 16:48  &885   &6.9      &61.06&2.87\\          
    11890 &2013.11.05 &X 3.3  & 22:36  &562   &4.8      &26.98&5.53\\
    11890 &2013.11.10 &X 1.1  & 05:36  &682   &2.3      &15.69&3.21\\
    12017 &2014.03.29 &X 1.0  & 18:12  &528   &5.0      &26.40&1.44\\
    12205 &2014.11.07 &X 1.6  & 18:08  &795   &12.0     &95.40&7.66\\
	12242 &2014.12.20 &X 1.8  & 01:25  &830   &22.0     &182.60&8.77\\         
    12297 &2015.03.11 &X 2.2  & 17:00  &240   &1.1      &2.64&6.18\\
    \hline

    \end{tabular}
  \end{table*}
\end{center}

\section{Statistical Results}
\label{sec:results}

{Following the same data reduction procedure as introduced in Section \ref{sec:observation}, we are able to plot the magnetic imprints as shown in Figures \ref{fig:map1}--\ref{fig:map3}. 

So far, many authors have given observational evidence of the type of magnetic imprint whose horizontal field irreversibly increased at the PIL (\citealt{Wangshuo2012a_2041-8205-745-2-L17}, \citealt{Wangshuo2012b_2041-8205-757-1-L5}; \citealt{Xudong2012_0004-637X-748-2-77}; \citealt{Petrie2012_0004-637X-759-1-50}, \citealt{Petrie2013}; \citealt{song2016relationship}), with a comprehensible physical interpretation (\citealt{Hudson2000_1538-4357-531-1-L75}; \citealt{Moore2001_0004-637X-552-2-833}; \citealt{Fisher2012}). However, for other types of magnetic imprints, more observational and theoretical discussions are needed before we can completely understand them.  In this sense, according to the spatial distribution of the ROI, more exactly, at PIL or not, we are able to categorize the magnetic imprints of {20 X-Class flares} into three types}:

\begin{enumerate}
\item Type \RN{1}: {$B_h$ increases at the PIL only, where there is a contiguous area of increasing $B_h$ that encompasses a PIL. In these events, we can identify a continuous area} with a strong enhancement of the horizontal field covering the PIL and the main flare region revealed by the 1600\ \AA \ AIA images, which is in agreement with previous results. Most of the events belong to this type.
\item Type \RN{2}: {$B_h$ increases both at the PIL and the penumbra, where there is a contiguous area of increasing $B_h$ that encompasses a PIL, as well as a contiguous area of increasing $B_h$ inside a penumbra. In a small number of events, besides the area at PIL, we can also identify a continuous area} with enhanced horizontal field at the penumbra, where the vertical magnetic field is greater than 1000 G. The events occurring in { AR 11520, AR 12158 and AR 12242 belong to this type.} 
\item {Type \RN{3}: $B_h$ increases in a separate and irregular way, spatially with little relation with the PIL. {Meanwhile, as shown in Figure \ref{fig:map3}, such changes also occur after the flare; so there is no temporal association with the flare.} Although we cannot exclude the relationship between the increases of $B_h$ and solar flares, the colored regions (with an increase of $B_h$ greater than 100 G) are identified as non-flare-related and mostly from noise or measurement errors. One X-class event in AR 11944 and three events in 12192 belong to this type.}
\end{enumerate}

Therefore, we reach the following statistical results: {(1) 16 out of 20 events embed clear magnetic imprints; 1 sample in AR 11944 and 3 samples in AR 12192 are exceptional (Type \RN{3}); (2) among the 16 flares with magnetic imprints, 13 events embed imprints only at the PIL as reported before, while 3 embed imprints both at the PIL and the sunspot penumbra;} {(3) the ROI in red boxes (Type \RN{1} and R1 in Type \RN{2}) spatially covered the flaring PIL, as revealed by the AIA data; the ROI in blue boxes (R2 in Type \RN{2}) are located on the flare ribbons.}

\begin{figure}[t!]
\label{fig:stat}
\centering
\includegraphics[width=0.45\textwidth]{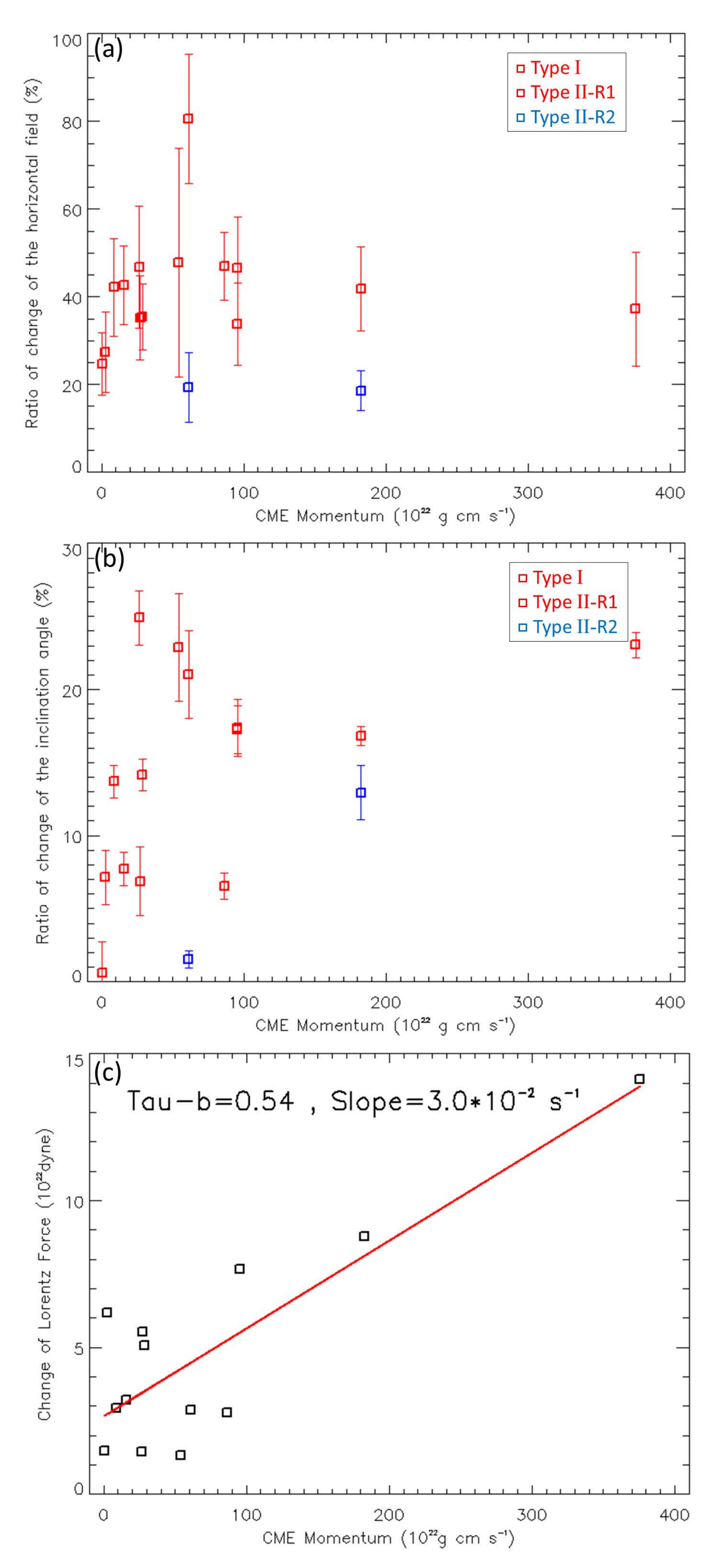}
\caption{(a) {Ratio of change of the horizontal field, with respect to the CME momentum. (b) Ratio of change of the inclination angle. {(c) Scatter plots of the CME momentum vs. the change of the Lorentz force, where the nonparametric Kendall's Tau-b coefficient is estimated to be 0.54. The red line shows the linear fitting to the data points, where the slope is estimated to be 3.0$\times10^{-2}$ $\mbox{s}^{-1}$, with the 90\% confidence interval being [2.3, 3.7] ($\times10^{-2}$ $\mbox{s}^{-1}$). The uncertainty of the slope is calculated by the bootstrap method with 50,000 iterations.}}}
\end{figure}

In addition to the above qualitative results, we further make a quantitative study on the change of the horizontal field, $\delta B_h$, and that of the inclination angle, $\delta \phi$, in all marked boxes for the {16 events} of Type \RN{1} and Type \RN{2}, as well as the parameters for the associated CME events. Note that, considering the poor relation between increase of horizontal field with solar flares, such quantitative statistics do not include the events in Type \RN{3}. We calculate the mean values of $B_h$, $\phi$ before the flare eruption\footnote{The mean value is calculated by the average of the measured values at five times instants immediately prior/posterior to the start/end time.}, denoted as $\langle \ B_h\ \rangle _{pre}$ and $\langle \ \phi \ \rangle _{pre}$, and the mean values of $B_h$ and $\phi$ after the eruption, denoted as $\langle \ B_h\ \rangle _{post}$, and $\langle \ \phi\ \rangle _{post}$. The difference values of them, $\langle \ \delta B_h\ \rangle$ and $\langle \ \delta \phi\ \rangle$, are listed in Table \ref{tab:flarelist}. {The errors in the table are calculated from error transfer formula}. {Since we are interested in the relationship between the magnetic imprints and the CME, we are able to plot the quantities describing the magnetic imprints in dependence of the CME momentum. The CME momentum of the flares in AR 11430, AR 11890 (2013.11.08) and AR 12158 are not available; so we do not show the three events in Figure \ref{fig:stat}}. As can be seen in Figure \ref{fig:stat}(a)--(b), the ratio of change of the horizontal field, $\frac{\langle \ B_h \ \rangle _{post} - \langle \  B_h \ \rangle _{pre}}{\langle \ B_h\ \rangle _{pre}}$, ranges from $\sim20\%$ to $\sim80\%$, with an average value of $\sim$40$\%$. Note that, given the size of error bars, except the events in AR 11166, 12297 and 11520 (R1), the ratio of change of the horizontal field at PIL (red boxes) tend to be a constant of $\sim$40$\%$, whereas the ratio of change of horizontal field at penumbra (blue boxes) tend to be $\sim$20$\%$. We suspect that these high values are due to the fact that all the flares in our sample are X-class ones. The change ratio should be different in other flares with lower classes. The ratio of change of the inclination angle, $\frac{\langle \ \phi \ \rangle _{post} - \langle \ \phi \ \rangle _{pre}}{\langle \ \phi\ \rangle _{pre}}$, is no greater than $25\%$. { No evident relationship can be seen between the CME momentum and the ratio of change of $B_h$ or $\phi$.}

{So far, there are some limitations in our method. Firstly, our method cannot completely remove the non-flare-associated change from the raw difference map. We visually exclude the non-flare-associated change area by excluding the $\delta B_h$ at post-flare stage and that at flare stage. However, inside the boxes, we cannot remove the non-flare-associated contribution of $\delta B_h$ from the ROI, which means that the magnetic field change in our statistics may still contain some non-flare-associated changes. Meanwhile, for the events whose magnetic imprints are not completely centralized at the PIL, some changes might be neglected outside the boxes. Secondly, similar to the definition of ROI in \cite{Wangshuo2012b_2041-8205-757-1-L5}, our temporal requirement of ROI that the horizontal magnetic field should increase over 100 G during the time interval from the flare onset to the end of flare end is empirical. It still remains to be tested whether this is always the case.}

The another objective of this paper is to test the relationship between the CME eruption and the change of the horizontal field. Considering the momentum conservation, it is natural to expect that an upward CME will result in a downward momentum impulse, which may lead to a change of magnetic field in the photosphere. According to the model of \cite{Fisher2012}, the momentum equation is:
\begin{equation}
\label{eq:momconservation}
M_{\mathrm{CME}}\ v=\frac{1}{2} \delta F_r\ \delta t,
\end{equation}
where $\delta F_r$ is the change of the Lorentz force acting on the photosphere (the absolute value\footnote{The change of Lorentz force acting on the outward atmosphere and that acting on the interior have the same magnitude, but opposite sign.} determined by Equation (\ref{eq:Lorentzforce})), $M_{\mathrm{CME}}$ is the CME mass and $v$ is the CME speed, both of which are available from the SOHO CME catalog\footnote{\href{$https://cdaw.gsfc.nasa.gov/CME_list/$}{\color{blue}https://cdaw.gsfc.nasa.gov/CME\_list/}}, and $\delta t$ is the reaction time of the physical process. When using this expression, we grossly assume that the entire mass of the CME moves with the same velocity, and that the work done against gravity is ignored. {Another assumption is made that the Lorentz force changes linearly during the time interval $\delta t$, which results in the factor $\frac{1}{2}$ in equation (\ref{eq:momconservation}). Since the theory behind the increase at the penumbra is unclear, we only study the regions of Type I and the R1 regions of Type II. Therefore, the value of ``back reaction'' time, $\delta t$, can be derived from the linear relationship between the CME momentum and the change of Lorentz force, $\delta F_r$, (see Figure \ref{fig:stat}(c)), whose slope equals to $\frac{2}{\delta t}$. The value of $\delta t$ of our samples is then calculated to be $\sim 70$ s, with the $90\%$ confidence interval being about 50 s to 90 s. Our estimation of back reaction time is longer than the rough estimation of 10 s given by \cite{Wangshuo2012b_2041-8205-757-1-L5}. One thing should be pointed out that, in reality, the Lorentz force may not change linearly. If the change is initially more rapid than a linear change, the factor in Equation (\ref{eq:momconservation}) will larger than $\frac{1}{2}$; if the change is initially slower than a linear change, the factor  will be smaller than $\frac{1}{2}$. This could explain the fairly large dispersion of the data points from the linear fitting in Figure \ref{fig:stat}(c).}

\section{Conclusions and Discussion}
\label{sec:conclusion}
After a statistical study of a sample of X-class flares with heliographic angle no greater than 45$\degree$ since SDO's launch, and further quantitative measurement of key parameters related to the magnetic field change, we reach the the main conclusions as follows.

First, most but not all the X-class solar flares would implement magnetic imprints at PIL on the photosphere. In our samples of {20 flares}, we observe that the horizontal field in one flare event in AR 11944 and three flare events in AR 12192 nearly remain unchanged at PIL during the flare time, similar to the behavior in the post-flare time (see Figure \ref{fig:map3}). {Our statistical study aims to explore the question presented by \cite{Xudong2017_0004-637X-839-1-67}: \textit{are the magnetic imprints universal?}} { The four aforementioned Type \RN{3} examples suggest that the magnetic imprints may not be universal in X-class flares, where categorization of the magnetic imprints is necessary. We still need to explore whether the universality exists in M-class or C-class flares.}

Second, we find that the magnetic imprints of X-class flares are located not only at the PIL (see \citealt{Wangshuo2012a_2041-8205-745-2-L17}; \citealt{song2016relationship}; \citealt{Xudong2017_0004-637X-839-1-67}) but also at the penumbra in some cases (see Region 2 in Figure \ref{fig:map2}(e)--(f)); or not at PIL (see Figure \ref{fig:map3}). This new finding needs to be checked with more observations and awaits a theoretical explanation. {We need to clarify how the change at the penumbra is related to the magnetic imprints that are usually found at the PIL, as it is not easily explained by the tether-cutting model \citep{Moore2001_0004-637X-552-2-833}.}

Third, we quantitatively evaluate the change of the horizontal field, inclination angle and the time duration of magnetic imprint process. As is seen in Table \ref{tab:flarelist} and Figure \ref{fig:stat}, It is found that the increase of the horizontal field ranges from ~150 G to ~450 G, and that the decrease of the inclination angle ranges from 1$\degree$ to 10$\degree$. {These  quantitative constrain the geometric change of the magnetic field} that has implications in constructing the physical model behind magnetic imprints. 

Fourth, we check the change of the Lorentz force on the photosphere in the ROI, which is proportional to integration of the $\delta B_h^2$. We find that this parameter is strongly correlated with the CME momentum (see Figure \ref{fig:stat}(d)). It suggests that a CME with a higher momentum is associated with a stronger magnetic imprint on the photosphere. In this sense, in the three X-flares in AR 12192, where there is no or very small CME momentum (also called as "confined flares" e.g. \citealt{Ji2003-APJL-1538-4357-595-2-L135}), it is natural that we cannot see intensive magnetic imprints. \cite{Sun2015-APJL-2041-8205-804-2-L28} interprets that the CMEs might be restricted by strong background fields. However, in the X-flare in AR 11944, we also cannot see the intensive magnetic imprint, regardless of very large momentum of the associated CME ($\sim$ 402.6$\times 10^{17}\mbox{g}\cdot \mbox{km}\cdot \mbox{s}^{-1}$, the highest among Table\ref{tab:flarelist}). This might be on account of its complex magnetic structure and unusual CME motion (\citealt{2015NatCo...6E7135M}; \citealt{Wang2015ApJ...814...80W}). { Moreover, we have estimated the back reaction time to be $\sim$ 70 s. As first introduced by \cite{Wangshuo2012b_2041-8205-757-1-L5}, since we can calculate $\delta F_r$ and measure the CME speed, $v$, once we know the key parameter back reaction time $\delta t$, we are able to estimate the CME mass. Based on the above analysis, we provide $\delta t \sim$ 70 s as a reference value for such CME mass estimation.} Here, the back reaction process is triggered by the magnetic reconnection, during which CME is propelled outward. The back reaction time, $\sim 70$ s, is shorter than the step-wise evolution time of the horizontal field on the photosphere, which is typically $\sim$48 minutes. Nevertheless, the physical nature of the reaction process is still unclear. In the future, we need to check the observational data with even higher cadence \citep{Xudong2017_0004-637X-839-1-67}, and perform MHD simulations for the eruptions, to understand the reaction process.

\acknowledgments
Acknowledgments: We are very grateful to the referee for constructive comments that helped improve the paper. We thank Kai Yang for help in data processing, and Xudong Sun for valuable comments on the manuscript. This project was supported by NSFC under grants 11733003, 11773016, 11703012, and 11533005, and NKBRSF under grant 2014CB744203.

\bibliographystyle{yahapj}
\bibliography{references}

\begin{thebibliography}{}
\providecommand\natexlab[1]{#1}
\providecommand\JournalTitle[1]{#1}

\bibitem[{Bobra {et~al.}(2014)Bobra, Sun, Hoeksema, Turmon, Liu, Hayashi,
  Barnes, \& Leka}]{Bobra2014}
Bobra, M.~G., Sun, X., Hoeksema, J.~T., {et~al.} 2014,
  \href{http://dx.doi.org/10.1007/s11207-014-0529-3}{\JournalTitle{Sol.Phys.},
  289, 3549}

\bibitem[{Fisher {et~al.}(2012)Fisher, Bercik, Welsch, \& Hudson}]{Fisher2012}
Fisher, G.~H., Bercik, D.~J., Welsch, B.~T., \& Hudson, H.~S. 2012,
  \href{http://dx.doi.org/10.1007/s11207-011-9907-2}{\JournalTitle{Sol.Phys.},
  277, 59}

\bibitem[{Hagenaar \& Shine(2005)}]{Hermance2005-0004-637X-635-1-659}
Hagenaar, H.~J., \& Shine, R.~A. 2005,
  \href{http://stacks.iop.org/0004-637X/635/i=1/a=659}{\JournalTitle{\apj},
  635, 659}

\bibitem[{{Hoeksema} {et~al.}(2014){Hoeksema}, {Liu}, {Hayashi}, {Sun},
  {Schou}, {Couvidat}, {Norton}, {Bobra}, {Centeno}, {Leka}, {Barnes}, \&
  {Turmon}}]{Hoeksema-2014SoPh..289.3483H}
{Hoeksema}, J.~T., {Liu}, Y., {Hayashi}, K., {et~al.} 2014,
  \href{http://dx.doi.org/10.1007/s11207-014-0516-8}{\JournalTitle{\solphys},
  289, 3483}

\bibitem[{{Hong} {et~al.}(2018){Hong}, {Ding}, {Li}, \&
  {Carlsson}}]{Hong-2018ApJ...857L...2H}
{Hong}, J., {Ding}, M.~D., {Li}, Y., \& {Carlsson}, M. 2018,
  \href{http://dx.doi.org/10.3847/2041-8213/aab9aa}{\JournalTitle{\apjl}, 857,
  L2}

\bibitem[{Hudson(2000)}]{Hudson2000_1538-4357-531-1-L75}
Hudson, H.~S. 2000,
  \href{http://stacks.iop.org/1538-4357/531/i=1/a=L75}{\JournalTitle{\apjl},
  531, L75}

\bibitem[{Iida {et~al.}(2012)Iida, Hagenaar, \&
  Yokoyama}]{Iida2012-0004-637X-752-2-149}
Iida, Y., Hagenaar, H.~J., \& Yokoyama, T. 2012,
  \href{http://stacks.iop.org/0004-637X/752/i=2/a=149}{\JournalTitle{\apj},
  752, 149}

\bibitem[{Ji {et~al.}(2003)Ji, Wang, Schmahl, Moon, \&
  Jiang}]{Ji2003-APJL-1538-4357-595-2-L135}
Ji, H., Wang, H., Schmahl, E.~J., Moon, Y.-J., \& Jiang, Y. 2003,
  \href{http://stacks.iop.org/1538-4357/595/i=2/a=L135}{\JournalTitle{\apjl},
  595, L135}

\bibitem[{Kopp \& Pneuman(1976)}]{Kopp1976}
Kopp, R.~A., \& Pneuman, G.~W. 1976,
  \href{http://dx.doi.org/10.1007/BF00206193}{\JournalTitle{Sol.Phys.}, 50, 85}

\bibitem[{Leka {et~al.}(2009)Leka, Barnes, Crouch, Metcalf, Gary, Jing, \&
  Liu}]{Leka2009}
Leka, K.~D., Barnes, G., Crouch, A.~D., {et~al.} 2009,
  \href{http://dx.doi.org/10.1007/s11207-009-9440-8}{\JournalTitle{Sol.Phys.},
  260, 83}

\bibitem[{Lemen {et~al.}(2012)Lemen, Title, Akin, Boerner, Chou, Drake, Duncan,
  Edwards, Friedlaender, Heyman, Hurlburt, Katz, Kushner, Levay, Lindgren,
  Mathur, McFeaters, Mitchell, Rehse, Schrijver, Springer, Stern, Tarbell,
  Wuelser, Wolfson, Yanari, Bookbinder, Cheimets, Caldwell, Deluca, Gates,
  Golub, Park, Podgorski, Bush, Scherrer, Gummin, Smith, Auker, Jerram, Pool,
  Soufli, Windt, Beardsley, Clapp, Lang, \& Waltham}]{Lemen2012}
Lemen, J.~R., Title, A.~M., Akin, D.~J., {et~al.} 2012,
  \href{http://dx.doi.org/10.1007/s11207-011-9776-8}{\JournalTitle{Sol.Phys.},
  275, 17}

\bibitem[{{Liu} {et~al.}(2015){Liu}, {Deng}, {Liu}, {Lee}, {Pariat},
  {Wiegelmann}, {Liu}, {Kleint}, \& {Wang}}]{Liu-C.-2015ApJ...812L..19L}
{Liu}, C., {Deng}, N., {Liu}, R., {et~al.} 2015,
  \href{http://dx.doi.org/10.1088/2041-8205/812/2/L19}{\JournalTitle{\apjl},
  812, L19}

\bibitem[{Liu {et~al.}(2012)Liu, Hoeksema, Scherrer, Schou, Couvidat, Bush,
  Duvall, Hayashi, Sun, \& Zhao}]{Liu2012}
Liu, Y., Hoeksema, J.~T., Scherrer, P.~H., {et~al.} 2012,
  \href{http://dx.doi.org/10.1007/s11207-012-9976-x}{\JournalTitle{Sol.Phys.},
  279, 295}

\bibitem[{{Masson} {et~al.}(2009){Masson}, {Pariat}, {Aulanier}, \&
  {Schrijver}}]{Masson-2009ApJ...700..559M}
{Masson}, S., {Pariat}, E., {Aulanier}, G., \& {Schrijver}, C.~J. 2009,
  \href{http://dx.doi.org/10.1088/0004-637X/700/1/559}{\JournalTitle{\apj},
  700, 559}

\bibitem[{{Masson} {et~al.}(2017){Masson}, {Pariat}, {Valori}, {Deng}, {Liu},
  {Wang}, \& {Reid}}]{Masson-2017A&A...604A..76M}
{Masson}, S., {Pariat}, {\'E}., {Valori}, G., {et~al.} 2017,
  \href{http://dx.doi.org/10.1051/0004-6361/201629654}{\JournalTitle{\aap},
  604, A76}

\bibitem[{Metcalf(1994)}]{Metcalf1994}
Metcalf, T.~R. 1994,
  \href{http://dx.doi.org/10.1007/BF00680593}{\JournalTitle{Sol.Phys.}, 155,
  235}

\bibitem[{Moore {et~al.}(2001)Moore, Sterling, Hudson, \&
  Lemen}]{Moore2001_0004-637X-552-2-833}
Moore, R.~L., Sterling, A.~C., Hudson, H.~S., \& Lemen, J.~R. 2001,
  \href{http://stacks.iop.org/0004-637X/552/i=2/a=833}{\JournalTitle{\apj},
  552, 833}

\bibitem[{{M{\"o}stl} {et~al.}(2015){M{\"o}stl}, {Rollett}, {Frahm}, {Liu},
  {Long}, {Colaninno}, {Reiss}, {Temmer}, {Farrugia}, {Posner}, {Dumbovi{\'c}},
  {Janvier}, {D{\'e}moulin}, {Boakes}, {Devos}, {Kraaikamp}, {Mays}, \&
  {Vr{\v{s}}nak}}]{2015NatCo...6E7135M}
{M{\"o}stl}, C., {Rollett}, T., {Frahm}, R.~A., {et~al.} 2015,
  \href{http://dx.doi.org/10.1038/ncomms8135}{\JournalTitle{Nature
  Communications}, 6, 7135}

\bibitem[{{Pesnell} {et~al.}(2012){Pesnell}, {Thompson}, \&
  {Chamberlin}}]{Pesnell-et-al-2012}
{Pesnell}, W.~D., {Thompson}, B.~J., \& {Chamberlin}, P.~C. 2012,
  \href{http://dx.doi.org/10.1007/s11207-011-9841-3}{\JournalTitle{\solphys},
  275, 3}

\bibitem[{Petrie(2012)}]{Petrie2012_0004-637X-759-1-50}
Petrie, G. J.~D. 2012,
  \href{http://stacks.iop.org/0004-637X/759/i=1/a=50}{\JournalTitle{\apj}, 759,
  50}

\bibitem[{Petrie(2013)}]{Petrie2013}
Petrie, G., J.~D. 2013,
  \href{http://dx.doi.org/10.1007/s11207-012-0071-0}{\JournalTitle{Sol.Phys.},
  287, 415}

\bibitem[{{Scherrer} {et~al.}(2012){Scherrer}, {Schou}, {Bush}, {Kosovichev},
  {Bogart}, {Hoeksema}, {Liu}, {Duvall}, {Zhao}, {Title}, {Schrijver},
  {Tarbell}, \& {Tomczyk}}]{Scherrer-et-al-2012}
{Scherrer}, P.~H., {Schou}, J., {Bush}, R.~I., {et~al.} 2012,
  \href{http://dx.doi.org/10.1007/s11207-011-9834-2}{\JournalTitle{\solphys},
  275, 207}

\bibitem[{Song \& Zhang(2016)}]{song2016relationship}
Song, Y., \& Zhang, M. 2016,
  \href{http://dx.doi.org/10.3847/0004-637X/826/2/173}{\JournalTitle{\apj},
  826, 173}

\bibitem[{Sudol \& Harvey(2005)}]{Sudol2005_0004-637X-635-1-647}
Sudol, J.~J., \& Harvey, J.~W. 2005,
  \href{http://stacks.iop.org/0004-637X/635/i=1/a=647}{\JournalTitle{\apj},
  635, 647}

\bibitem[{Sun {et~al.}(2017)Sun, Hoeksema, Liu, Kazachenko, \&
  Chen}]{Xudong2017_0004-637X-839-1-67}
Sun, X., Hoeksema, J.~T., Liu, Y., Kazachenko, M., \& Chen, R. 2017,
  \href{http://stacks.iop.org/0004-637X/839/i=1/a=67}{\JournalTitle{\apj}, 839,
  67}

\bibitem[{Sun {et~al.}(2012)Sun, Hoeksema, Liu, Wiegelmann, Hayashi, Chen, \&
  Thalmann}]{Xudong2012_0004-637X-748-2-77}
Sun, X., Hoeksema, J.~T., Liu, Y., {et~al.} 2012,
  \href{http://stacks.iop.org/0004-637X/748/i=2/a=77}{\JournalTitle{\apj}, 748,
  77}

\bibitem[{Sun {et~al.}(2015)Sun, Bobra, Hoeksema, Liu, Li, Shen, Couvidat,
  Norton, \& Fisher}]{Sun2015-APJL-2041-8205-804-2-L28}
Sun, X., Bobra, M.~G., Hoeksema, J.~T., {et~al.} 2015,
  \href{http://stacks.iop.org/2041-8205/804/i=2/a=L28}{\JournalTitle{\apjl},
  804, L28}

\bibitem[{Wang(1992)}]{Wang1992}
Wang, H. 1992,
  \href{http://dx.doi.org/10.1007/BF00148431}{\JournalTitle{Sol.Phys.}, 140,
  85}

\bibitem[{{Wang} {et~al.}(1994){Wang}, {Ewell}, {Zirin}, \&
  {Ai}}]{1994ApJ...424..436W}
{Wang}, H., {Ewell}, Jr., M.~W., {Zirin}, H., \& {Ai}, G. 1994,
  \href{http://dx.doi.org/10.1086/173901}{\JournalTitle{\apj}, 424, 436}

\bibitem[{{Wang} \& {Liu}(2012)}]{Wang-Liu-2012ApJ-circular-ribbon-flares}
{Wang}, H., \& {Liu}, C. 2012,
  \href{http://dx.doi.org/10.1088/0004-637X/760/2/101}{\JournalTitle{\apj},
  760, 101}

\bibitem[{Wang {et~al.}(2002)Wang, Spirock, Qiu, Ji, Yurchyshyn, Moon, Denker,
  \& Goode}]{Wang2002_0004-637X-576-1-497}
Wang, H., Spirock, T.~J., Qiu, J., {et~al.} 2002,
  \href{http://stacks.iop.org/0004-637X/576/i=1/a=497}{\JournalTitle{\apj},
  576, 497}

\bibitem[{{Wang} {et~al.}(2015){Wang}, {Liu}, {Dai}, {Yang}, {Huang}, \&
  {Hu}}]{Wang2015ApJ...814...80W}
{Wang}, R., {Liu}, Y.~D., {Dai}, X., {et~al.} 2015,
  \href{http://dx.doi.org/10.1088/0004-637X/814/1/80}{\JournalTitle{\apj}, 814,
  80}

\bibitem[{Wang {et~al.}(2012{\natexlab{a}})Wang, Liu, Liu, Deng, Liu, \&
  Wang}]{Wangshuo2012a_2041-8205-745-2-L17}
Wang, S., Liu, C., Liu, R., {et~al.} 2012{\natexlab{a}},
  \href{http://stacks.iop.org/2041-8205/745/i=2/a=L17}{\JournalTitle{\apjl},
  745, L17}

\bibitem[{Wang {et~al.}(2012{\natexlab{b}})Wang, Liu, \&
  Wang}]{Wangshuo2012b_2041-8205-757-1-L5}
Wang, S., Liu, C., \& Wang, H. 2012{\natexlab{b}},
  \href{http://stacks.iop.org/2041-8205/757/i=1/a=L5}{\JournalTitle{\apjl},
  757, L5}

\bibitem[{{Yang} {et~al.}(2015){Yang}, {Guo}, \&
  {Ding}}]{Yang-2015ApJ...806..171Y}
{Yang}, K., {Guo}, Y., \& {Ding}, M.~D. 2015,
  \href{http://dx.doi.org/10.1088/0004-637X/806/2/171}{\JournalTitle{\apj},
  806, 171}

\bibitem[{{Ye} {et~al.}(2016){Ye}, {Liu}, \& {Wang}}]{ye2016irreversible}
{Ye}, D.-D., {Liu}, C., \& {Wang}, H. 2016,
  \href{http://dx.doi.org/10.1088/1674-4527/16/6/095}{\JournalTitle{RAA}, 16,
  95}

\end{thebibliography}

\end{document}